\documentclass[iop]{emulateapj}
\usepackage{apjfonts}
\newcommand{\Msun}{\,M_{\sun}}
\newcommand{\Mmin}{M_{\mathrm{min}}}
\newcommand{\Mmax}{M_{\mathrm{max}}}
\newcommand{\Msph}{{M_{*}}}
\newcommand{\Mpc}{\,\mathrm{Mpc}}
\newcommand{\kpc}{\,\mathrm{kpc}}
\newcommand{\pc}{\,\mathrm{pc}}
\newcommand{\yr}{\,\mathrm{yr}}
\newcommand{\Gyr}{\,\mathrm{Gyr}}
\newcommand{\kms}{\mathrm{km\, s^{-1}}}
\newcommand{\Sersic}{S\'ersic\ }

\begin{document}

\slugcomment{Submitted to ApJ}
\shortauthors{GNEDIN ET AL.}
\shorttitle{}

\title{Co-Evolution of Galactic Nuclei and Globular Cluster Systems}

\author{ Oleg Y. Gnedin\altaffilmark{1},
         Jeremiah P. Ostriker\altaffilmark{2,3},
         Scott Tremaine\altaffilmark{4}
}

\altaffiltext{1}{University of Michigan, Department of Astronomy, 
   Ann Arbor, MI 48109; ognedin@umich.edu}
\altaffiltext{2}{Princeton University Observatory,
   Princeton, NJ 08544}
\altaffiltext{3}{Columbia University, Department of Astronomy \& Astrophysics,
   New York, NY 10027}
\altaffiltext{4}{Institute for Advanced Study,
   Princeton, NJ 08540}

\begin{abstract}
  We revisit the hypothesis that dense galactic nuclei are formed from
  inspiraling globular clusters.  Recent advances in understanding of
  the continuous formation of globular clusters over cosmic time and
  the concurrent evolution of the galaxy stellar distribution allow us
  to construct a simple model that matches the observed spatial and
  mass distributions of clusters in the Galaxy and the giant
  elliptical galaxy M87.  In order to compare with observations, we
  model the effects of dynamical friction and dynamical evolution,
  including stellar mass loss, tidal stripping of stars, and tidal
  disruption of clusters by the growing galactic nucleus.  We find
  that inspiraling globular clusters form a dense central structure,
  with mass and radius comparable to the typical values in observed
  nuclear star clusters (NSCs) in late-type and low-mass early-type
  galaxies.  The density contrast associated with the NSC is less
  pronounced in giant elliptical galaxies.  Our results indicate that
  the NSC mass as a fraction of mass of the galaxy stellar spheroid
  scales as $M_{NSC}/\Msph \approx 0.0025 \, \Msph_{,11}^{-0.5}$.
  Thus disrupted globular clusters could contribute most of the mass
  of NSCs in galaxies with stellar mass below $10^{11}\Msun$.  The
  inner part of the accumulated cluster may seed the growth of a
  central black hole via stellar dynamical core collapse, thereby
  relieving the problem of how to form luminous quasars at high
  redshift.  The seed black hole may reach $\sim 10^5\Msun$ within
  $\la 1\Gyr$ of the beginning of globular cluster formation.
\end{abstract}


\section{Introduction}

The centers of most luminous galaxies contain supermassive black holes
(BH) and/or nuclear star clusters (NSC).  The history of NSC formation
and BH growth can involve a variety of different physical processes.
Some of the central stars may have formed {\it in situ}, while others
may have formed elsewhere and been brought to the center.  In this
paper we explore the simplest version of the latter hypothesis: that
due to dynamical friction, globular star clusters have spiraled from
larger radii to the inner few parsecs of a galaxy and there deposited
a dense and massive concentration of stars.  These stars would form a
NSC which, via various dynamical processes, may produce the seed of a
supermassive black hole.  The timescale for cluster infall and NSC
growth, set by the processes of dynamical friction and concurrent
tidal disruption, is unrelated to the timescale of accretion growth of
the black hole.  This difference in the timescales may have important
implications for the formation of the first supermassive black holes.

We do not presume the {\it in situ} and {\it ex situ} scenarios of NSC
formation to be mutually exclusive.  The presence of a NSC and/or a BH
will attract gas to the central region and induce star formation.
This seems to be the case in our own galactic center, where the bulk
of the stars are billions of years old but the presence of very
massive stars indicates recent star formation.

The idea that globular cluster orbits may decay to the galaxy center
by dynamical friction is not new \citep{tremaine_etal75}.  In a series
of papers, \citet{capuzzo93}, \citet{capuzzo_miocchi08},
\citet{antonini_etal12}, and \citet{antonini13} have explored cluster
inspiral and disruption leading to the formation of a NSC.  Major
uncertainties in this analysis are whether there is a pre-existing
black hole at the galaxy center and what fraction of the incoming
cluster mass, if any, is eventually added to the black hole.  Assuming
a pre-existing black hole of $4\times 10^{6}\Msun$ at the center of a
stellar system resembling the Galactic bulge, \citet{antonini_etal12}
find that the inspiral of globular clusters could build a significant
NSC.  Although the observed cluster at the Galactic center contains
many young stars that must have formed {\it in situ}, over half of the
cluster could be old ($\sim 10\Gyr$) stars acquired from disrupted
globular clusters.

This conclusion is consistent with the analysis of the kinematics of
two dwarf NSCs by \citet{hartmann_etal11}; they find that cluster
merging can explain most of the kinematic and spectroscopic properties
of the NSCs, but that comparable {\it in-situ\,} star formation is
also required.  However, the fraction of {\it in-situ\,} stars may
increase with the brightness of the NSC: absorption-line spectroscopy
of dwarf elliptical galaxies in the Virgo cluster by
\citet{paudel_etal11} indicates that stars in brighter nuclei have
younger ages and higher metallicity than could be contributed by old
globular clusters.  This trend is also supported by the HST surveys of
galaxies in the Virgo and Fornax clusters \citep{turner_etal12}.
%
In spiral galaxies, nuclei tend to be smaller and younger in late-type
galaxies relative to early-type spirals \citep{rossa_etal06}.
%
\citet{leigh_etal12} have recently revisited the scaling relations of
NSCs and argued that the {\it present-day} distribution of globular
clusters is insufficient to build up the observed nuclei by dynamical
friction and disruption.  However, they leave open the possibility
that the globular cluster systems may have contained more
massive and more numerous clusters in the past, which could provide the required
stellar material.  In fact, the deficit in globular cluster numbers in
the central part of our Galaxy, and other galaxies, as compared to the
stellar distribution, can be taken as evidence that dynamical
processes have depleted the globular cluster system
\citep[e.g.,][]{lotz_etal01, capuzzo_mastrobuono09}.

A comprehensive theoretical study of NSC formation by inspiraling
clusters was done by \citet{agarwal_milosa11}.  They assumed that all
stars in a galaxy form instantaneously in star clusters with a
power-law mass distribution having a slope $\beta=2$ between $\Mmin$
and $\Mmax$ (cf. eq.~\ref{eq:beta} below).  As massive clusters
spiral towards the center, they form a NSC.  The steep power-law
slope of 2 implied that most of the accumulated mass is from the most
massive clusters, and therefore the mass in the inner few $\pc$ and the
prominence of the NSC depend strongly on the adopted (and largely,
unconstrained) value of $\Mmax$.

\citet{antonini13} further extended that model and explored how the
expected NSC properties depend on galaxy mass.  He found that the
formation of NSCs is suppressed in giant galaxies relative to dwarf
galaxies.  The absolute normalization of the cluster density profile
still remained uncertain because of the unknown cluster formation rate
and maximum cluster mass.

Since the mass and spatial distribution of globular clusters are well
studied in nearby galaxies, they provide additional, and critically
important, constraints on $\Mmax$ and thus on the whole scenario of
NSC formation from clusters.  In this work we set up a similar
calculation to \citet{agarwal_milosa11} and \citet{antonini13} but
choose the parameters to match the observed globular cluster profile
in two well-studied galaxies: the Milky Way (MW) and the giant
elliptical M87.  We also make additional improvements: we account for
the observed cosmological evolution of the galaxy stellar profile,
model the continuous addition of star clusters over time, and include
mass loss due to stellar evolution.

We begin by building a simple model for the formation and evolution of
globular cluster systems (\S\ref{sec:model}); a central assumption of
this model is that the initial spatial distributions of the clusters
and field stars are the same.  We then apply this model to the Galaxy,
where the most detailed observations are available, in \S\ref{sec:mw}.
We extend the model to the M87 system in \S\ref{sec:m87} and find
that, in order to reproduce the observed distribution of globular
clusters, we must account for the evolution of the galaxy structure
over cosmic time, in a fashion that matches both recent observational
data and cosmological simulations.  Once the evolved models match the
basic current observed properties of the globular cluster system, we
investigate the accumulation of stars stripped from the star clusters
at the galaxy center.  In \S\ref{sec:lowermass} we apply this model to
scaled-down versions of M87 to investigate how the NSC mass depends on
the host galaxy mass.  Finally, in \S\ref{sec:discussion} we discuss
whether the accumulated stellar distribution could undergo collisional
evolution and core collapse into a massive black hole rapidly enough
to alleviate the timescale problem for high-redshift quasars.

\section{A simple model for the formation and evolution of globular
  cluster systems}
  \label{sec:model}

\subsection{Formation}

In massive early-type galaxies at the present time, the mass of the
globular cluster system is roughly proportional to the total stellar
mass, $M_{\mathrm{GC}} \sim (1-5)\times 10^{-3}\, M_{*}$
\citep[e.g.,][]{spitler_forbes09, georgiev_etal10, harris_etal13}.
Guided by this relation, we adopt the simple ansatz that in a given
galaxy the cluster formation rate was a fixed fraction
$f_{\mathrm{GC},i}$ of the overall star formation rate:
\begin{equation}
  {dM_{GC} \over dt} = f_{\mathrm{GC},i} \; {dM_{*} \over dt}.
\end{equation}
Specifically, we assume that the globular cluster system forms
continuously between redshifts $z_{i}=6$ and $z_{f}=0$, with the
initial density distribution of clusters proportional to the density
distribution of field stars formed at the same epoch.  This is the
simplest assumption; in detail, the cluster fraction may scale with
the local gas density \citep[e.g.,][]{kruijssen12}.  Our results are
insensitive to the particular choice of $z_{i}$, but considering
redshifts higher than 6 requires severe extrapolation of the currently
available observational scaling relations.  By ``formation'' we
include both the {\it in situ\,} star and cluster
formation, and the accretion of stars and clusters formed in satellite
galaxies that later merge with the central galaxy.

To account for the dynamical disruption of globular clusters over
time, we choose a larger normalization of the cluster formation rate
($f_{\mathrm{GC},i}$) than the present fraction of stellar mass locked
in globular clusters, $f_{\mathrm{GC},f} = (1-5)\times 10^{-3}$.
We set $f_{\mathrm{GC},i}$ by matching the final number of clusters in
our models for the MW and M87 systems to the observed number.

The masses of individual clusters are drawn from the power-law
distribution
\begin{equation}
   dN/dM \propto M^{-\beta}, \quad \Mmin < M < \Mmax,
\label{eq:beta}
\end{equation}
that is observed for young massive star clusters in nearby galaxies
\citep[e.g.,][]{zhang_fall99, gieles09, larsen09, chandar_etal10,
chandar_etal10b, chandar_etal11}.  We adopt the average slope
$\beta=2$, and also explore the effect of varying this slope between
1.8 and 2.2, as allowed by observational uncertainties.

The lower limit of the mass spectrum is set to $\Mmin = 10^{4}\Msun$.
Most of our results are insensitive to this choice, since we expect
all $10^{4}\Msun$ clusters (and even most $10^{5}\Msun$ clusters) to
be dynamically disrupted by the present time.

In contrast, the upper limit of the mass spectrum affects both the
shape of the cluster mass function at zero redshift and the total
amount of cluster debris accumulated at the center.  If we let $\Mmax$
increase to infinity, the mass function would be too broad in most
galaxies and the central mass concentration would be too large.  By
matching the observed mass function, we constrain this important upper
limit.

We describe below several galaxy models, appropriate for the Galaxy,
the giant elliptical galaxy M87, and scaled-down variants of M87.
For the Galaxy, the upper limit of the cluster mass
spectrum is taken to be $\Mmax = 10^{7}\Msun$, while for M87 we
increase the upper limit to $2\times 10^{7}\Msun$, to account for its
broader cluster mass function.

\subsection{Disruption}

As soon as the clusters have formed, they begin to lose mass via
stellar winds and the dynamical ejection of stars through two-body
relaxation and stripping by the galactic tidal field.

We model the stellar mass loss using the main-sequence lifetimes from
\citet{hurley_etal00} and the stellar remnant masses from
\citet{chernoff_weinberg90}.  The time-dependent mass loss rate is
calculated as in \citet{prieto_gnedin08}.  Over 10 Gyr, stellar
evolution reduces the cluster mass by up to $f_{\mathrm{se}} = 40\%$
for a \citet{chabrier03} stellar IMF.

We model the dynamical evaporation of stars in the tidal field of the
host galaxy, but not tidal shocks (due, for example, to pericenter
passages close to the galactic center) because we do not calculate
individual cluster orbits.  Effectively, we assume a circular
trajectory and take the radius $r$ to be the time-averaged radius of a
true, likely eccentric, cluster orbit. A recent estimate of the
disruption time based on direct $N$-body simulations is given by
\citet{gieles_baumgardt08}.  We write it as
\begin{eqnarray}
 {dM \over dt} & = & -{M \over \min{(t_{\mathrm{tid}}, t_{\mathrm{iso}})}},
   \label{eq:dis}\\
 t_{\mathrm{tid}}(r,M) & \approx & 10\Gyr \left[{M(t) \over 2\times 10^{5}\Msun}\right]^{\alpha} P(r),
   \label{eq:ttid}\\
 P(r) & \equiv & 41.4 \left({r \over \kpc}\right) 
         \left({V_{c}(r) \over \kms}\right)^{-1}.\nonumber
\end{eqnarray}
Here $V_{c}(r)$ is the circular velocity of the galaxy at a distance
$r$ from its center.  The factor $P(r)$ is effectively a normalized
period of rotation around the galaxy, which reflects the strength of
the local tidal field, which in turn controls the rate of evaporation.
Models with $\alpha=1$ and $P=1$ can reproduce the observed cluster
mass function in the Galaxy \citep[e.g.,][]{fall_zhang01,
prieto_gnedin08}.  However, more recent $N$-body calculations suggest
$\alpha \approx 2/3$ \citep[e.g.,][]{baumgardt01, gieles_baumgardt08}
because of the lingering of stripped stars near the tidal radius, and
we adopt this lower value for our models.

The formula for $t_{\mathrm{tid}}(r,M)$ in equation (\ref{eq:dis}) is
derived from $N$-body models in which the tidal field is strong.
However, in the limit of a weak tidal field (which occurs, for
example, in the outer galaxy) the evaporation is mostly controlled
internally and not by the tidal radius.  To account for this, we
calculate the evaporation time in isolation as a multiple of the
half-mass relaxation time, $t_{rh}$:
\begin{equation}
  t_{\mathrm{iso}}(M) = {t_{rh} \over 2.5\, \xi_{e}} 
    \approx 17\Gyr \left[{M(t) \over 2\times 10^{5}\Msun}\right],
  \label{eq:tiso}
\end{equation}
where $\xi_{e} = 0.0074$ is the \citet{ambartsumian38} rate for a
single stellar mass model, and the factor 2.5 corrects it for
multi-mass models \citep[e.g.,][]{gieles_etal11}.  

We then take the disruption time for a given cluster to be the smaller
of $t_{\mathrm{tid}}$ and $t_{\mathrm{iso}}$.  The isolated
evaporation is more important for clusters at large radii and helps to
reduce the spatial gradient of the cluster mass function.  The weak
tidal field regime may be especially relevant for giant galaxies, as
recent HST observations of \citet{brodie_etal11} indicate that many
globular clusters in M87 may not be tidally limited.

When a cluster arrives in the immediate vicinity of the galactic
center, the tidal forces may be strong enough to immediately tear
apart the whole cluster.  This happens when the stellar density at a
characteristic place in the cluster, such as the core or half-mass
radius, falls below the ambient density
\citep[e.g.,][]{antonini_etal12}.  Here we adopt the average density
at the half-mass radius as such a density: $\rho_{h} \equiv
\frac{1}{2} M / (\frac{4\pi}{3} R_{h}^{3})$.  To estimate the ambient
density in the inner galaxy, $\rho_{*}$, dominated by stars, we note
that the adopted field stellar mass, as well as the calculated
mass of the NSC, are increasing with radius roughly as $M_{*}(r) \propto
r^{2}$, which gives the average density $\rho_{*}(r) \approx
M_{*}(r)/(2\pi r^{3})$.  Rewriting this in terms of the circular
velocity, we obtain a criterion for direct cluster disruption if
\begin{equation}
  \rho_{h} < {V_{c}(r)^{2} \over 2\pi G\, r^{2}}.
  \label{eq:dirdis}
\end{equation}
The observed half-mass density of Galactic globular clusters has a
wide dispersion, with the median value $\rho_{h} \sim
10^{3}\Msun\pc^{-3}$.  The density of an individual cluster may depend
on its evolutionary stage and location in the galaxy.  According to
\citet{gieles_etal11}, star clusters form with a very compact size and
then expand until they reach their tidal radius.  The expansion phase
lasts longer for the more massive clusters than for the less massive
ones.  During this phase, cluster density scales as $\rho_{h} \propto
M^{2}$.  In later stages of cluster evolution the median density
changes slowly and the core of a massive cluster can remain very
dense, even when the cluster has expanded to the tidal radius and its
outer stars are stripped.  Thus it is the median cluster density that
determines how long massive clusters could withstand extremely strong
tidal forces near the center.

As the NSC begins to build up, its stellar density will exceed even
the high density of infalling globular clusters and these clusters
will be directly disrupted before reaching the galaxy center.  We
continuously update the values of $V_{c}(r)$ to include the
accumulated stellar and gaseous debris from the disrupted clusters.
We deposit the whole remaining cluster mass at the radius where
Equation~(\ref{eq:dirdis}) is first satisfied.  To check for this
condition, we set a fixed value of $\rho_{h} = 10^{3}\Msun\pc^{-3}$
for low-mass clusters ($M < 10^{5}\Msun$).  For more massive clusters,
the density is assumed to vary with mass as $\rho_{h} = 10^{3}\,
(M/10^{5}\Msun)^{2}\Msun\pc^{-3}$.  In the most massive clusters, we
limit $\rho_{h}$ to a maximum of $10^{5}\Msun\pc^{-3}$, as it is about
the highest observed half-mass density of a stellar cluster.  The
cluster mass $M$ here is the current value of the mass before
disruption, not the initial mass.

\subsection{Dynamical Friction}
  \label{sec:df}

We include the effect of dynamical friction on cluster orbits by
gradually reducing the radius $r$ as suggested by
\citet{binney_tremaine08}, equation (8.24):
\begin{equation}
  {dr^{2} \over dt} = -{r^{2} \over t_{\rm df}(r,M)},
  \label{eq:df}
\end{equation}
\begin{equation}
  t_{\rm df} = 0.45\Gyr \left({r \over \kpc}\right)^{2}
                \left({V_{c}(r) \over \kms}\right) 
                \left({M(t) \over 10^{5}\Msun}\right)^{-1} f_{\epsilon},
  \label{eq:tdf}
\end{equation}
where we assume a Coulomb logarithm $\ln\Lambda=5.8$ and take $V_{c}
= \sqrt{2}\, \sigma$.  This expression ignores many uncertain factors,
such as the details of stellar velocity distribution and possible
variation of the Coulomb logarithm with cluster mass.  Our simplified
expression suffices for the present calculation, where we also
approximate the galaxy stellar mass profile by a single \Sersic model.

We include one important correction for eccentricity of cluster
orbits, $f_{\epsilon}$.  It is usually parameterized as a function of
the ratio of the orbital angular momentum to its maximum value for a
given energy, $J/J_{c}(E)$.  Results quoted in the literature range from
$f_{\epsilon} \approx (J/J_{c})^{0.78}$ \citep{lacey_cole93} to
$f_{\epsilon} \approx (J/J_{c})^{0.53}$ \citep{vandenbosch_etal99} to
$f_{\epsilon} \approx (J/J_{c})^{0.4}$ \citep{colpi_etal99}.  More recent
analysis of N-body simulations of the inspiral of a satellite galaxy
within its host galaxy by \citet{boylan-kolchin_etal08} found instead
$f_{\epsilon} \approx \exp{[1.9\, (J/J_{c} - 1)]}$.  The differences of
the fitting expressions are due to the increasing complexity of the
simulations and inclusion of the effects of tidal mass loss of the
satellite along its trajectory.

Since we do not know the initial distribution of globular cluster
eccentricities, we take as an estimate the distribution of orbit
circularity of satellite halos in cosmological simulations of galaxy
cluster formation by \citet{jiang_etal08}.  That distribution peaks
around $J/J_{c}(E) \approx 0.5$.  Substituting this peak value into
the various expressions above, we obtain a range of possible values of
the correction factor $f_{\epsilon} \approx 0.39-0.76$.  In order to
keep our model simple, we adopt a single value in the middle of this
range, $f_{\epsilon} = 0.5$, for all our calculations.  However, it is
important to keep in mind the additional uncertainty associated with
this factor.

The differential equations for tidal mass loss, stellar mass loss, and
dynamical friction are integrated simultaneously.  The time step is
limited to be 0.02 times the smaller of the current disruption time or
current dynamical friction time (we checked that taking a smaller
tolerance of 0.01 produces indistinguishable results).  We trace the
gradual deposition of mass by stripped stars and stellar winds along
the cluster path towards the center, and apply dynamical friction only
to the remaining bound part of the cluster.  Most of the mass loss due
to stellar winds happens early in the cluster evolution and that mass
is deposited close to the initial location of the cluster.  Most of
the mass deposited near the galactic center is in the form of stars.

To convert redshift into cosmic time we use the WMAP7 cosmological
model, with $\Omega_{m} = 0.272$, $\Omega_{\Lambda} = 0.728$, $h =
0.704$.  This relation is essentially unchanged with the latest Planck
satellite measurements.

\begin{figure}[t]
\vspace{-0.4cm}
\centerline{\epsfxsize=3.5in \epsffile{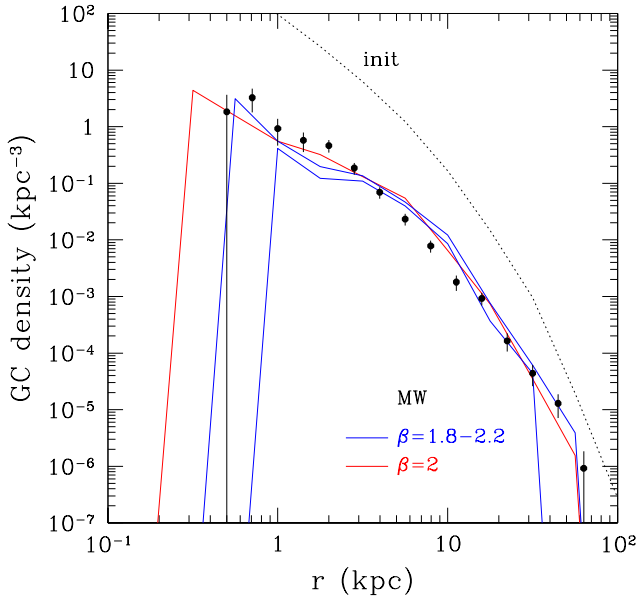}}
\vspace{-0.4cm}
\caption{Number density of Galactic globular clusters surviving to the
  present time, in the fiducial model (red solid line) and its
  variants with a different slope $\beta$ (eq.\ \ref{eq:beta}) of the
  initial cluster mass function (blue lines).  The dotted line shows
  the initial distribution of all model clusters at redshift
  $z_{i}=3$; most of these are low-mass clusters that evaporate
  completely.  Symbols with error bars show the observed number
  density of Galactic globular clusters \citep[from][]{harris96}.}
  \label{fig:den_mw}
\vspace{0.2cm}
\end{figure}

\begin{table*}[t]
\begin{center}
\caption{}
\label{tab:sim}
\begin{tabular}{lcllcccccrc}
\tableline\tableline\\
\multicolumn{1}{l}{System} &
\multicolumn{1}{c}{$M_{*}\,(\Msun)$} &
\multicolumn{1}{c}{$f_{\mathrm{GC},i}$} &
\multicolumn{1}{c}{$f_{\mathrm{GC},f}$} &
\multicolumn{1}{c}{$M_{10}\,(\Msun)$} &
\multicolumn{1}{c}{$t_{1/2}\,(\Gyr)$} &
\multicolumn{1}{c}{$R_{\mathrm{h}}\,(\pc)$} &
\multicolumn{1}{c}{$R_{\mathrm{h,proj}}\,(\pc)$} &
\multicolumn{1}{c}{$r_{\mathrm{cc}}\,(\pc)$} &
\multicolumn{1}{c}{$M_{\mathrm{BH,exp}}\,(\Msun)$} &
\multicolumn{1}{c}{$M_{\mathrm{BH,obs}}\,(\Msun)$}
\\[2mm] \tableline\\
MW  & $5\times 10^{10}$ & 0.012 & 0.0012 & $4.7\times 10^{7}$ & 0.68 & 2.8 & 2.1 & 9.5 & $(2-5)\times 10^{4}$ & $4\times 10^{6}$ \\
EG2 & $5\times 10^{10}$ & 0.04  & 0.0014 & $1.6\times 10^{8}$ & 0.20 & 5.5 & 5.1 & 7.0 & $(0.3-2)\times 10^{5}$ & ($2\times 10^{8}$)$^a$\\
EG1 & $2\times 10^{11}$ & 0.04  & 0.0034 & $5.0\times 10^{8}$ & 1.11 & 7.1 & 5.3 & 5.5 & $(1-6)\times 10^{5}$ & ($9\times 10^{8}$)$^a$\\
M87 & $8\times 10^{11}$ & 0.04  & 0.0054 & $5.8\times 10^{8}$ & 1.38 & 7.4 & 5.7 & 5.3 & $(1-7)\times 10^{5}$ & $(3.5-6)\times 10^{9}$ \\[2mm]
\tableline
\end{tabular}
\end{center}
\vspace{0cm} {\bf Columns:} $M_{*}$ -- galaxy stellar mass;
$f_{\mathrm{GC},i}$, $f_{\mathrm{GC},f}$ -- initial and final cluster
mass fractions; $M_{10}$ -- stellar mass accumulated within inner
$10\pc$; $t_{1/2}$ -- time to reach half of the final mass within
$10\pc$; $R_{\mathrm{h}}$ -- half-mass radius of NSC;
$R_{\mathrm{h,proj}}$ -- projected half-mass radius of NSC;
$r_{\mathrm{cc}}$ -- radius of core-collapsed region at $z=0$;
$M_{\mathrm{BH,exp}}$ -- expected black hole mass after core
  collapse, equation~(\ref{eq:bhexp}); 
$M_{\mathrm{BH}}$ -- observed (or inferred) central black
  hole mass.\\ 
$^a$ The inferred black hole mass is from the scaling
  relation of \citet{gultekin_etal09}.
\vspace{0.3cm}
\end{table*}

\section{Evolution of the globular cluster system in the Galaxy}
  \label{sec:mw}

We approximate the density distribution of stars in the Galaxy as a
spherical \Sersic profile with total mass $M_{*} = 5\times
10^{10}\Msun$, effective radius $R_{e} = 4\kpc$, and concentration
index $n_{s} = 2.2$.  We also include a dark-matter halo with an NFW
profile having mass $M_{h} = 10^{12}\Msun$ and scale radius $r_{s} =
20\kpc$.  Our simple mass model satisfies basic observational
constraints from \citet{launhardt_etal02} within $230\pc$, from
\citet{mcmillan_binney10} at $8\kpc$, and from \citet{gnedin_etal10}
out to $80\kpc$ from the Galactic center.  While we ignore the
distinction between the bulge and the disk, the single stellar mass
profile should suffice for our calculation of cluster inspiral and
disruption because most clusters deposited near the center formed in
the inner $1\kpc$, where the bulge dominates.  
We initialize the clusters in a spherical distribution, proportional
to that of the field stars, and treat the position $r$ as the
time-averaged radius of an assumed orbit, as discussed in
\S\ref{sec:df}.  As a starting point, we take this mass model to be
fixed in time from the moment the globular clusters form.

Unlike the elliptical galaxy models described below, for the MW model
we choose a single formation epoch for all clusters.  We assume that
clusters formed at $z_{i}=3$ and calculate their evolution for
$11.5\Gyr$ until the present.  This assumption is partly justified
because most of the observed Galactic clusters are old and metal-poor.
The adopted value of $z_{i}$ should therefore be interpreted as the
epoch of the peak of globular cluster formation rate.

Based on the adopted stellar profile, we generate the initial
positions of about 8000 model clusters (the exact number depends on the
slope $\beta$), corresponding to the initial cluster fraction
$f_{\mathrm{GC},i} = 0.012$.  We construct a fiducial model with $\beta=2$
and $\Mmax = 10^{7}\Msun$, and variants with $\beta=1.8$ and
$2.2$ for the same $\Mmax$, and with $\Mmax = 5\times 10^{6}\Msun$
and $2\times 10^{7}\Msun$ for the same $\beta=2$.  

Figure~\ref{fig:den_mw} shows the initial and final ($z_{f}=0$) radial
distributions of the model clusters.  Over 97\% of the initial
clusters are disrupted, but this wholesale destruction is primarily
the effect of including low-mass clusters in the initial distribution:
all clusters with $M \la 10^{5}\Msun$ are disrupted, and therefore the
reduction in numbers is dominated by clusters in the range
$10^{4}$--$10^{5}\Msun$.  More relevant is the consistency of the
predicted final distribution with the observed one \citep[from][the
updated 2010 edition is at
http://physwww.mcmaster.ca/$\sim$harris/mwgc.dat]{harris96}.  The
excellent agreement seen in Figure~\ref{fig:den_mw} persists for all
considered values of $\beta=1.8$--$2.2$ and $\Mmax = 5\times 10^{6} -
2\times 10^{7}\Msun$.  It is also important to note that the
flattening of the observed density profile near the center may be
partially due to the distance measurement errors and dust obscuration.

We chose the initial cluster mass fraction to match the normalization
of the cluster profile at large radii.  Alternatively,
\citet{bonatto_bica12} considered several scenarios for the evolution
of the cluster mass function and found the best fit for the initial
mass of all globular clusters after the gas removal phase
($10^{7}\yr$) of $M_{GC,i} \sim 3\times 10^{8}\Msun$.  Our initial
cluster mass is a factor of two higher, but that includes also the
mass that will be lost by stellar evolution in the first $10^{7}\yr$.
The rest of the difference could be attributed to
\citeauthor{bonatto_bica12}'s choice of lower stellar mass-to-light
ratio.

\begin{figure}[t]
\vspace{-0.4cm}
\centerline{\epsfxsize=3.5in \epsffile{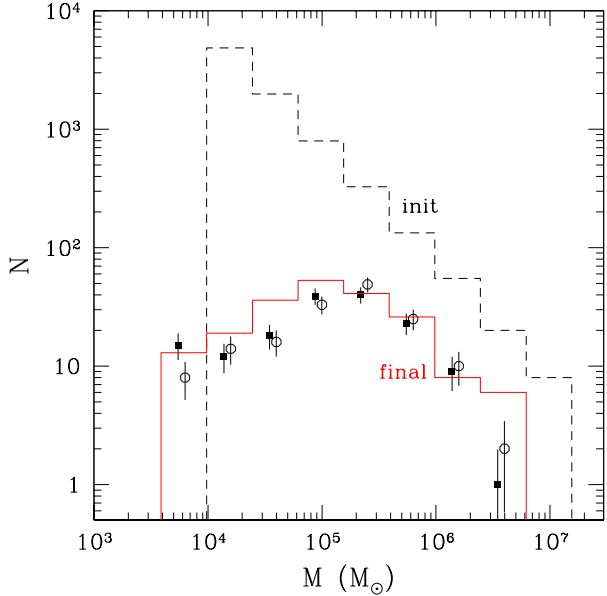}}
\vspace{-0.4cm}
\caption{Initial (dashed histogram) and evolved (solid red histogram)
  cluster mass function in the fiducial Milky Way model.  Symbols show
  the observed mass function of Galactic globular clusters, for two
  choices of the mass-to-light ratio: constant $M/L_V =
  3\Msun/L_{\sun}$ (open circles), and $M/L_V$ increasing gradually
  with cluster luminosity (filled squares) as derived using multi-mass
  King models by \citet{mandushev_etal91}.  Error-bars are from
  Poisson counting statistics.}
  \label{fig:mf_mw}
\vspace{0.2cm}
\end{figure}

In Figure~\ref{fig:mf_mw} we show the transformation of the cluster
mass function from the initial power law to a roughly log-normal
distribution at $z_{f}=0$, in satisfactory agreement with the
observations.  Most of the low-mass clusters have evaporated, and some
of the high-mass clusters have spiraled into the center.  In order to
convert the observed luminosity of Galactic globular clusters into
mass, we use two choices for the mass-to-light ratio.  In the first,
we take a constant $M/L_V = 3\Msun/L_{\sun}$, as frequently used in
the literature for an old metal-poor stellar population.  In the
second, we take a variable $M/L_V$, motivated by available dynamical
measurements that show $M/L_V$ increasing systematically with cluster
luminosity, albeit with large scatter.  Most measurements and models
\citep[e.g.,][]{mclaughlin_vandermarel05, kruijssen_portegies09} are
based on single-mass King (or similar) profiles, which may
underestimate the mass-to-light ratio by up to a factor of two,
compared to more realistic multi-mass King models, as shown by
\citet{mandushev_etal91}.  We adopt the average results of the
multi-mass modeling of \citet{mandushev_etal91}, which can be written
as $M/L_V \approx 0.85\, (L_V/L_{\sun})^{0.08}$ in solar units.
\citet{kruijssen_mieske09} showed that part of the luminosity
dependence arises from preferential escape of low mass stars over the
course of dynamical evolution, which is faster in low-mass clusters.

\begin{figure}[t]
\vspace{-0.4cm}
\centerline{\epsfxsize=3.5in \epsffile{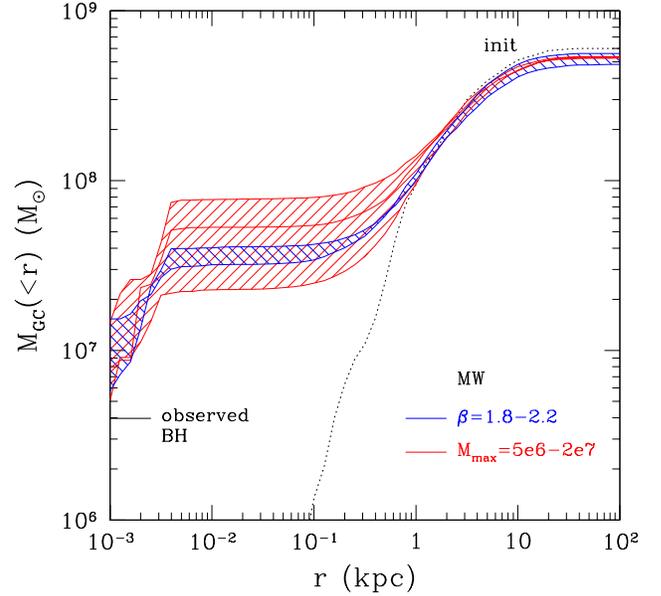}}
\vspace{-0.4cm}
\caption{Stellar mass deposited by disrupted clusters in our models of
  cluster evolution in the Galaxy.  The dotted line shows the
  cumulative mass distribution with radius of the initial cluster
  population.  The blue shaded region shows the cumulative mass of
  stars from disrupted clusters for models in which the slope of the
  initial cluster mass function varies in the range $\beta = 1.8-2.2$.
  The red shaded region shows the same quantity when the maximum
  cluster mass varies in the range $\Mmax= 5\times 10^{6} - 2\times
  10^{7}\Msun$.  Larger $\Mmax$ results in larger accumulated mass.
  The observed central black-hole mass is marked at the bottom left
  part of the plot.}
  \label{fig:cl_mw}
\vspace{0.2cm}
\end{figure}

Figure~\ref{fig:mf_mw} also shows that taking $\Mmax > 10^{7}\Msun$
would significantly overproduce the number of high-mass
clusters.  On the other hand, the case of $\Mmax= 5\times 10^{6}\Msun$
has no clusters with current mass above $2\times 10^{6}\Msun$.  Any
lower $\Mmax$ would result in a disagreement with observations.  Thus
our choice of $\Mmax = 10^{7}\Msun$ appears to be appropriate for the
Galaxy model.

Different slopes of the initial cluster mass function also produce
noticeable changes in the final distribution.  The shallow slope
$\beta=1.8$ skews the mass function towards high masses, resulting in
8 clusters with $M > 3\times 10^{6}\Msun$.  The steep slope $\beta =
2.2$ pushes in the other direction, lowering the mean cluster mass
below $10^{5}\Msun$.  The fiducial choice $\beta = 2$ again matches
the data best.

Despite the radial dependence of the evaporation time (equation
\ref{eq:ttid}), the peak of the mass function does not show systematic
variation with galactocentric radius.  The gradient may be difficult
to detect because of the low number of surviving clusters in the MW
model.  The peak mass remains in the range $(1-2)\times 10^{5}\Msun$
out to 50~kpc from the center.

Having found a simple model that is consistent with the observed
radial distribution and mass function of the Galactic globular
clusters, we now calculate the mass of stars deposited by the
disrupted clusters (Figure~\ref{fig:cl_mw}).  We discount the mass
lost via stellar winds, assuming that this is incorporated in the
interstellar medium of the Galaxy.  Most of the mass loss from winds
takes place when the clusters are young and far from the Galactic
center.  We also passively evolve the deposited stars to the present,
so that the plotted stellar mass is as it would be observed.

Dynamical friction causes massive clusters in the central kpc or so to
spiral towards the center. For example, Figure~\ref{fig:cl_mw}
demonstrates that clusters with a combined mass of $4\times
10^{7}\Msun$ sink to within a few pc of the center.  Very close to the
center, typically within $10\pc$, the clusters completely disrupt so
formally the mass does not reach radius $r=0$.  Instead, the stripped
stars form a dense stellar core (which we show in
Figure~\ref{fig:dend_all}) whose tidal field in turn disrupts new
infalling clusters.

Figure~\ref{fig:cl_mw} shows that the mass accumulated within a few pc
of the Galactic center exceeds the observed mass of the central black
hole by a factor of 2--20, depending on the slope $\beta$ and
especially on the maximum cluster mass, $\Mmax$.  More massive
clusters spiral in more efficiently, and thus dominate the mass of
the stellar nucleus.

Our conclusion that a significant stellar mass can accumulate within
the inner few parsecs is similar to that of \citet{antonini_etal12},
who found that a stream of disrupted massive globular clusters would
form a dense core of 1--2 pc in size.  They show that such a nuclear
cluster would subsequently shrink in size due to gravitational
interactions between the stars (``collisional evolution''), to a core
radius of about $0.5\pc$.

Thus we can draw a robust conclusion that the stellar mass accumulated
at the Galactic center is several times larger than the observed
black-hole mass.  We will investigate the likely state of this
stellar nucleus later in~\S\ref{sec:discussion}.

\section{Evolution of the globular cluster system in M87}
  \label{sec:m87}

We assume that the distribution of stars in M87 follows a \Sersic
profile with mass $M_{*} = 8\times 10^{11}\Msun$, effective radius
$R_{e} = 30\kpc$, and concentration index $n_{s} \simeq 8$; this fit
to the density profile holds over the range $0.7-35\kpc$
\citep{kormendy_etal09}.  In the innermost kpc the profile flattens to
$n_{s} \approx 3$.  However, to obtain an upper limit on the disrupted
cluster mass, we adopt a single profile with $n_{s} = 8$ at all radii.
This increases the inner stellar mass by a factor of two within
$1\kpc$, but only by 8\% within $10\kpc$.  We compare the observed
profile with our model at the end of Section~\ref{sec:m87results}.

We also include a dark-matter halo with an NFW profile having mass
$M_{h} = 2.7\times 10^{13}\Msun$ and scale radius $r_{s} = 50\kpc$.
This simple two-component model provides an accurate fit to the
overall mass distribution obtained by \citet{gebhardt_thomas09} up to
$100\kpc$.

Based on the adopted stellar profile, we generate the initial
positions of $\sim 400,000$ model clusters, corresponding to an
initial cluster fraction $f_{\mathrm{GC},i} = 0.04$.  This is 3--4
times higher than the assumed fraction in the Galaxy, because M87
(like other brightest cluster galaxies) is known to contain
proportionately more globular clusters \citep{harris01}.  In order to
match its broader cluster mass function \citep[e.g.,][]{kundu_etal99,
waters_etal06}, we increase the maximum allowed cluster mass to
$M_{\mathrm{max}} = 2\times 10^{7}\Msun$.

\begin{figure}[t]
\vspace{-0.4cm} \centerline{\epsfxsize=3.5in \epsffile{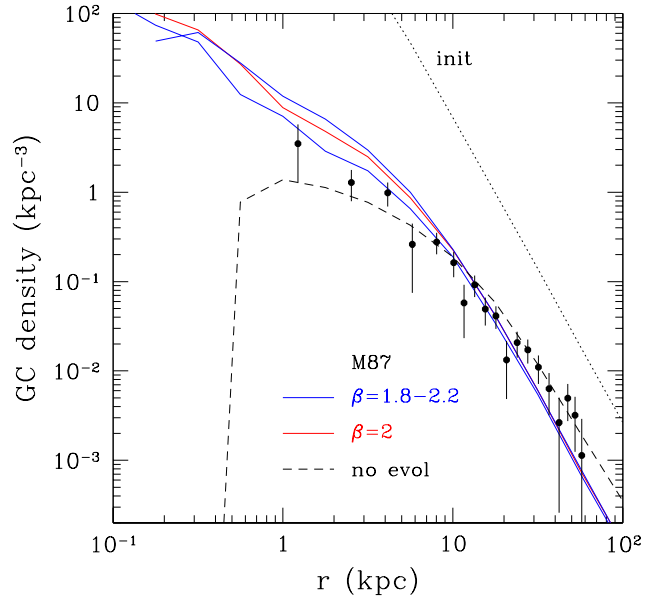}}
\vspace{-0.4cm}
\caption{Number density versus radius for the M87 clusters in the
  fiducial model without evolution (dashed line) and with evolution
  (solid red line), along with its variants with different slope
  $\beta$ (solid blue lines).  Symbols with error bars show the
  observed number density of clusters in M87.}
  \label{fig:den_m87}
\vspace{0.2cm}
\end{figure}

Figure~\ref{fig:den_m87} shows the initial ($z_{i}=3$) and final
($z_{f}=0$) density profiles of the model clusters (dashed line), as
well as the observed density profile.  The latter was deconvolved from
the surface number density by \citet{mclaughlin99} and
\citet{tamura_etal06}.  The zero-redshift model is consistent with the
observed profile at large radii but underestimates it in the inner
several kpc.

A likely explanation for this discrepancy comes from the observation
that elliptical galaxies evolve with redshift, in stellar mass, size,
and surface-brightness profile. To take this effect into account, we
incorporate the evolution of the galaxy stellar profile and the
extended formation of star clusters over the whole cosmic history of
the galaxy.  In addition, we model the accretion of satellite galaxies
as well as their associated black holes and globular cluster systems,
as we describe below.

\subsection{Evolution of the host galaxy}
  \label{sec:evo1}

We model the evolution of the stellar distribution of elliptical
galaxies like M87 using the parametrization of
\citet{vandokkum_etal10}, which is based on a stacked sample of giant
elliptical galaxies in the redshift range $0.2 < z < 2.2$.  The sample
is chosen such that the number density of galaxies ($2\times
10^{-4}\Mpc^{-3}$) is the same at all considered redshifts, and
therefore the evolution of mass should reflect the actual growth of
the same galaxies, not the changing sample.  This method ignores the
changes in galaxy numbers due to merging, which are expected to be
less significant for these massive galaxies at $z \la 2$, when most of
the mass accretion is via minor mergers
\citep[e.g.,][]{hopkins_etal10, oser_etal12, leja_etal13,
  behroozi_etal13b}.  Note that we did not include such evolution in
our model of the Milky Way clusters because the relevant data are
lacking.

Fitting the photometry of these galaxies to \Sersic profiles, van
Dokkum et al.\ find that the evolution of the total stellar mass,
effective radius, and \Sersic index are:
\begin{eqnarray}
  M_{*}(z) & = & M_{*}(0) \; (1+z)^{-0.67},\nonumber\\
  R_{e}(z) & = & R_{e}(0) \; (1+z)^{-1.27},\label{eq:ms}\\
  n_{s}(z) & = & n_{s}(0) \; (1+z)^{-0.95}.\nonumber
\end{eqnarray}
The observations constrain these relations only for $z < 2.2$, but by
necessity we extrapolate these relations to higher redshift, which
adds uncertainty to our results.  Recently, \citet{nipoti_etal12} have
derived the evolution of structural parameters of a larger sample of
elliptical galaxies in the redshift range $z=1.3-2.3$, using three
different models for the halo mass-stellar mass relation.  Our adopted
power-law slope for the redshift evolution of stellar mass falls
within the range of the slopes ($-0.6$ to $-1.5$) derived by Nipoti et
al.  At low redshift, the increase of the effective radius is
consistent with an independent measurement for brightest cluster
galaxies, by a factor $1.70 \pm 0.15$ from $z \approx 0.5$ to $z
\approx 0$ \citep{ascaso_etal11}.  The adopted evolution is also
similar to the build-up of stellar mass in the recent numerical
simulations of elliptical galaxy formation by \citet{naab_etal09} and
\citet{oser_etal12}.

Although we allow the stellar mass profile to evolve with redshift, we
keep the halo mass profile fixed, for several reasons: (i) the inner
region of the dark-matter halo has probably stabilized by $z \approx
3$; (ii) the stellar mass dominates over the dark matter mass in the
inner 6 kpc, so the exact profile of the inner halo is unimportant;
(iii) the region where globular clusters experience significant
dynamical friction is even smaller, $\la 1\kpc$ (see
Figure~\ref{fig:cl_m87}), and hence even more strongly dominated by
the stellar mass.

We use the evolving stellar profile to calculate the circular velocity
$V_{c}(r,t)$ that enters the disruption time and dynamical friction
calculation (equations \ref{eq:dis} and \ref{eq:df}).

\subsection{Accretion of satellite galaxies}
  \label{sec:evo2}

The growth in stellar mass over cosmic time described by equation
(\ref{eq:ms}) can be crudely divided into two channels: stars that
form in the main body of the elliptical galaxy ({\it in situ\,}
growth) and stars that are brought in by accreting satellite galaxies
(``accretion'' growth).  The exact breakdown between these two
channels is not known. However, \citet{vandokkum_etal10} estimate that
at $z \ga 1-2$ the central galaxy produces the majority of stars,
while at lower redshifts satellite accretion dominates the stellar
mass build-up.  Similar conclusions were reached by
\citet{oser_etal10} by analyzing numerical simulations of galaxy
formation.

\begin{figure}[t]
\vspace{-0.2cm}
\centerline{\epsfxsize=3.5in \epsffile{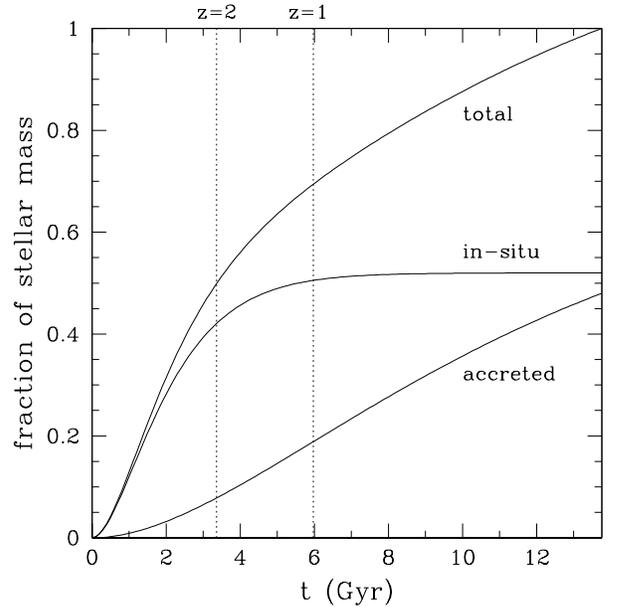}}
\vspace{-0.4cm}
\caption{Growth of stellar mass in an average elliptical galaxy, based
  on the analysis of \citet{vandokkum_etal10}.  The plot shows the
  fraction of the $z=0$ mass accumulated at a given cosmic time,
  measured from the Big Bang.  The {\it in-situ\,} stars are those
  formed in the main galactic system, typically within $\sim 10\kpc$
  from the center.  The ``accreted'' stars are brought in by mergers
  and disruption of satellite galaxies, typically from larger
  distances.}
  \label{fig:mass_growth}
\vspace{0.2cm}
\end{figure}

We have approximated the relative contribution from each channel by
fitting the observed trends in \citet{vandokkum_etal10} to growth
rates of the form $dM_{*}/dt \propto (t/\tau)\ e^{-t/\tau}$, where time
$t$ is measured from the Big Bang. We find $\tau \approx 1.1\Gyr$ for {\it
in-situ\,} growth and $\tau \approx 6\Gyr$ for accretion growth.
Figure~\ref{fig:mass_growth} shows the fractional contribution of each
channel to the total stellar mass.

\subsection{Continuous formation of clusters}
  \label{sec:evo3}

Given the evolution of the overall stellar component, we allow for
continuous (rather than instantaneous) formation of star clusters.  In
order to consider the earliest formation of clusters, we extend the
initial epoch to $z_{i}=6$.  The details and exact timing of
globular cluster formation are still poorly understood
and it may be desirable to consider even earlier epochs, however
further extrapolation of the scaling relations (Equation~\ref{eq:ms})
is extremely uncertain.

We split the cosmic time between $z_{i}$ and $z_{f}=0$ into 20
equal-duration intervals (of 640~Myr) and form a group of clusters in
each interval.  The total mass of the group formed in each interval is
proportional to the increase of stellar mass within that interval:
$\Delta M_{GC} = f_{\mathrm{GC},i} \; \Delta M_{*}$, with the same
factor $f_{\mathrm{GC},i}$ as before.  This algorithm produces the
same number of clusters as in the instantaneous model, but now their
formation is distributed over the Hubble time.  In each interval, the
new clusters are added according to the spatial profile of the field
stars at that epoch.

Some of the clusters in each group form {\it in situ\,} and some come
from accreted satellites.  Since the accretion events are not actual
star formation but the addition of already existing stars, we assign a
formation epoch for the accreted clusters that differs from the time
of accretion.  Specifically, we draw an epoch prior to accretion, from
the distribution of the satellite's star formation history, assuming
it is the same as for the central galaxy
(Figure~\ref{fig:mass_growth}).  The dynamical evolution of the
satellite clusters prior to accretion is calculated simply using
Equation~(\ref{eq:tiso}), thus ignoring tidal truncation.  The initial
location of the satellite clusters is also not important, as they are
added later to the central galaxy along with the other satellite
stars.  After accretion the satellite clusters are subject to the same
dynamical friction and evolution as the {\it in-situ\,} clusters.

We also ignore potential migration of old clusters as the host
galaxy's stellar mass grows.  If the existing clusters were on
circular orbits and conserved their angular momentum, their orbital
radius would shrink as the mass interior to the orbit grows.  On the
other hand, mergers that lead to the increase of the stellar effective
radius should excite the orbits of the old clusters to larger radii.
It is not clear where the sum of the two effects would point.  In view
of these uncertainties, we do not adjust the positions of the old
clusters as the galaxy grows, and only include the monotonic effect of
dynamical friction.

\begin{figure}[t]
\vspace{-0.4cm}
\centerline{\epsfxsize=3.5in \epsffile{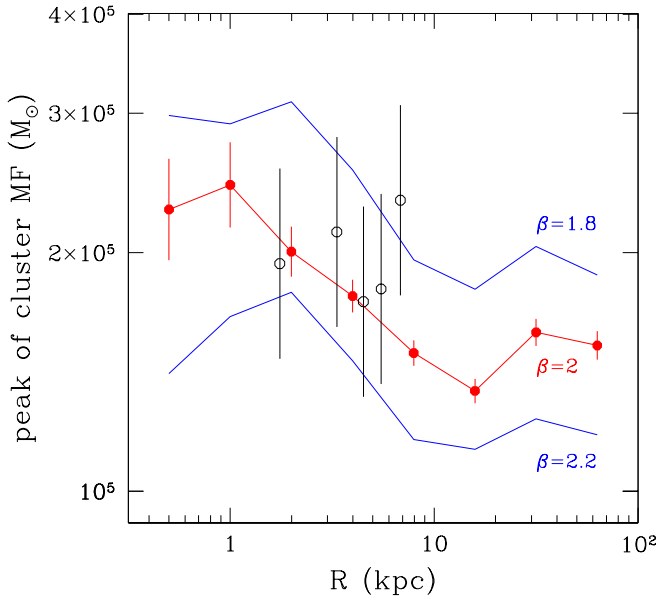}}
\vspace{-0.4cm}
\caption{Peak of the projected mass function of clusters in the M87
  system (solid red symbols with error bars), as a function of
  cylindrical distance from the galaxy center.  Blue lines show the
  possible range due to the variation of the slope of the initial
  cluster mass function $\beta$ between 1.8 and 2.2.  Open black
  circles with errorbars show the observed peaks from
  \citet{kundu_etal99} and their $1\sigma$ errors.}
  \label{fig:mfr_m87}
\vspace{0.2cm}
\end{figure}

\subsection{Evolution of the cluster mass function}

Many studies of the dynamical evolution of globular clusters have
demonstrated that the cluster mass function develops a characteristic
peaked shape as a result of preferential disruption of low-mass
clusters \citep[see, e.g., review by][]{portegies_etal10}.  Our model
similarly transforms the initial power-law mass function into a peaked
distribution, well fit by a log-normal function.  We have already
shown this in the case of the Galaxy (Figure~\ref{fig:mf_mw}).  For
M87, the peak mass $\log{M}_{\rm peak} \approx 5.2$ matches the
observations in the inner regions \citep{waters_etal06}.  However, the
model mass function is somewhat broader than the observations, with a
log-normal standard deviation $\sigma_{M} = 0.75$~dex compared to the
observed value $\sigma_{M} = 0.57$~dex for clusters located between 1
and $9\kpc$ of the galaxy center \citep{kundu_etal99}.

We have also checked that the case of $M_{\mathrm{max}} =
10^{7}\Msun$, as in the Galaxy model, results in no clusters with
current mass above $4\times 10^{6}\Msun$, whereas the data require at
least a dozen of more massive clusters.  This justifies raising the
maximum cluster mass in the M87 model to at least $2\times
10^{7}\Msun$.

Faster disruption of low-mass clusters in the inner parts of the
galaxy leads to an increase of the average mass of surviving clusters
as we approach the center.  To compare this prediction for M87 with
the HST observations of \citet{kundu_etal99}, we have randomly
projected cluster positions in our model and calculated the peak of
the cluster mass function in bins of cylindrical radius $R$.
Figure~\ref{fig:mfr_m87} shows the gradient of the peak mass in the
model and the absence of such a gradient in the observations, but the
observational errors are large enough that the predictions are still
statistically consistent with the data.

\subsection{Accretion of satellite black holes}
  \label{sec:evo4}

Most massive galaxies in the local universe appear to contain a
massive black hole at their center.  The satellites accreted by our
central galaxy are also likely to bring with them their black hole.
We add these satellite black holes to the host galaxy at the time of
accretion and then follow their inspiral towards the center.

For simplicity, we attribute all the accreted mass in a given 640~Myr
interval to a single satellite, which would contribute one black hole.
We estimate the mass of the latter using the observed relation between
black-hole mass and host galaxy luminosity at $z \approx 0$
\citep{gultekin_etal09}:
\begin{equation}
  M_{\rm bh} \approx 9\times 10^{8}\Msun 
     \left({L_{V} \over 10^{11}\, L_{\sun}}\right)^{1.1}
     f_{\mathrm{bh}}(z).
\end{equation}
Assuming a stellar mass-to-light ratio $M_{*}/L_{V} = 2
\Msun/L_{\sun}$, we obtain $M_{\rm bh,sat}/M_{*,\rm sat} \approx
0.004$ in the relevant mass range ($M_{*} \ga 5\times 10 ^{10}\Msun$).
This ratio would be correspondingly lower if we had chosen a higher
value of $M_{*}/L_{V}$.  Current observational constraints from
\citet{sani_etal11} give the black hole mass fraction between 0.001
and 0.004, although \citet{kormendy_ho13} and \citet{graham_scott13}
have recently advocated that this should be revised upward to about
0.005.  Our choice of $M_{*}/L_{V}$ is therefore consistent with the
observed range.

Black holes added at redshift $z>0$ would have smaller masses because
they have had less time to grow.  We account for this effect by the
factor $0 \le f_{\mathrm{bh}}(z) \le 1$, taken to be the fraction of
the overall cosmic black hole density attained at the accretion epoch,
using Figure~1 from \citet{yu_tremaine02}.  This fraction is
$f_{\mathrm{bh}} \approx 0$ at $z>6$, and it rises monotonically to
$f_{\mathrm{bh}}=1$ at $z=0$.

Our algorithm gives the maximum contribution of satellite black holes.
If we were to split the accreted mass in several satellites, it would
slightly reduce the combined black hole mass at the center of the main
galaxy, because dynamical friction is less efficient for smaller black
holes.

In order to determine where to place a satellite black hole at the
time of accretion, we calculate the radius where the enclosed added
stellar mass is 0.004 of the total added stellar mass.  This implies that
these black holes are deposited along with the innermost satellite
stars, which are likely to be gravitationally bound to them.  We then
follow the orbital evolution of the satellite black holes due to
dynamical friction, again including the factor $f_{\epsilon} = 0.5$
describing orbit eccentricity, until they sink to the galactic center.

The details of satellite black hole orbits do not affect the stellar
mass accumulated at the center from disrupted globular clusters.  We
have checked that the accumulated mass is not visibly changed even if
we move the black holes to the center as soon as the satellites are
accreted onto the host galaxy.

\begin{figure}[t]
\vspace{-0.4cm}
\centerline{\epsfxsize=3.5in \epsffile{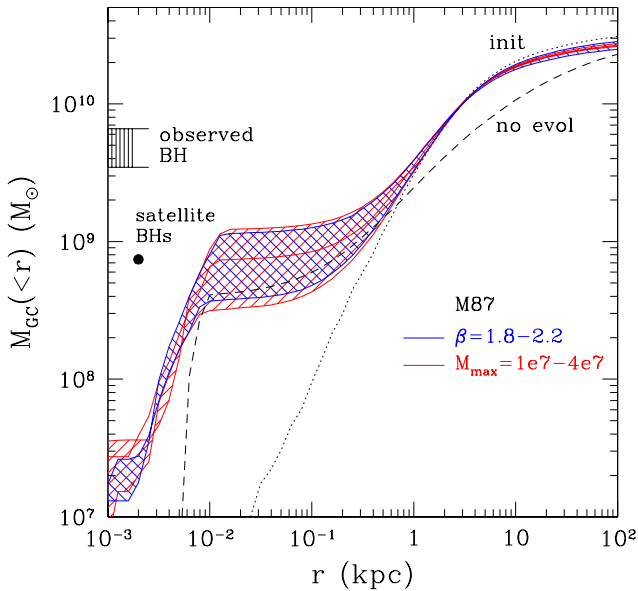}}
\vspace{-0.4cm}
\caption{Stellar mass deposited by disrupted globular clusters in our
  M87 models.  The line notation is as in Figure~\ref{fig:cl_mw}.  The
  red shaded region corresponds to the range $\Mmax =
  10^{7} - 4\times 10^{7}\Msun$.  The dashed line shows the case
  without evolution of the galaxy, while the solid lines include
  evolution.  The filled circle shows the sum of the masses of all
  black holes in satellite galaxies brought in to the center by
  dynamical friction.  The observed black hole mass measurements
  differ for the stellar-dynamics and gas-dynamic methods and are
  shown as a shaded range.}
  \label{fig:cl_m87}
\vspace{0.2cm}
\end{figure}

\subsection{Results}
  \label{sec:m87results}

We now discuss the results from the model for the evolution of M87,
its globular clusters, and its satellites described in \S\S
\ref{sec:evo1}--\ref{sec:evo4}.  First consider
Figure~\ref{fig:den_m87}, showing the radial distribution of the
globular clusters at the present time. The dashed line, from the model
with no evolution of the galaxy, underestimates the observed cluster
density profile, as we noted just before the start of
\S\ref{sec:evo1}. After including evolution, the model now
overestimates the density in the inner $\sim 2\kpc$.  At large radii
the model cluster profile now falls systematically below the observed
profile but is still consistent with it within the errors.  Thus the
inclusion and exclusion of the evolution brackets the actual cluster
distribution, which suffices for the purpose of our models.

Figure~\ref{fig:cl_m87} shows the cumulative mass of disrupted
clusters in the M87 models that include galaxy evolution.  The stellar
mass accumulated within 10 pc of the center is a factor of ten below
the black-hole mass measurement from stellar dynamics by
\citet{gebhardt_etal11}: $M_{\rm bh,M87} = (6.6 \pm 0.4)\times
10^{9}\Msun$.  However, the discrepancy is less with the black-hole
mass based on gas-dynamical measurements \citep{walsh_etal13}: $M_{\rm
bh,M87} = (3.5^{+0.9}_{-0.7})\times 10^9\Msun$.

In the model without evolution, the accumulated mass is even lower (by
40\% in the fiducial model).  
Within 1 pc, the accumulated mass falls by another order of magnitude
in the model with evolution, and essentially vanishes without
evolution.

The plot also shows the cumulative mass of the satellite BHs that have
sunk to the center by dynamical friction.  These contribute a mass
that is comparable to the mass that accumulates in the center from
globular clusters, but still is insufficient to provide the majority
of the observed black-hole mass.

The sharp drop of the accumulated mass profile at $r < 10\pc$ is due
to the formation of a dense NSC.  The tide from the NSC disrupts
globular clusters as they approach the center; roughly speaking, as
the NSC accumulates mass it grows in radius so that its mean density
remains comparable to the mean density of the clusters it destroys.
In the inner few parsecs the space density of the NSC reaches $\sim
2\times 10^{5}\Msun\pc^{-3}$, while the projected surface density
reaches $\sim 5\times 10^{6}\Msun\pc^{-2}$.

\begin{figure}[t]
\vspace{-0.4cm}
\centerline{\epsfxsize=3.5in \epsffile{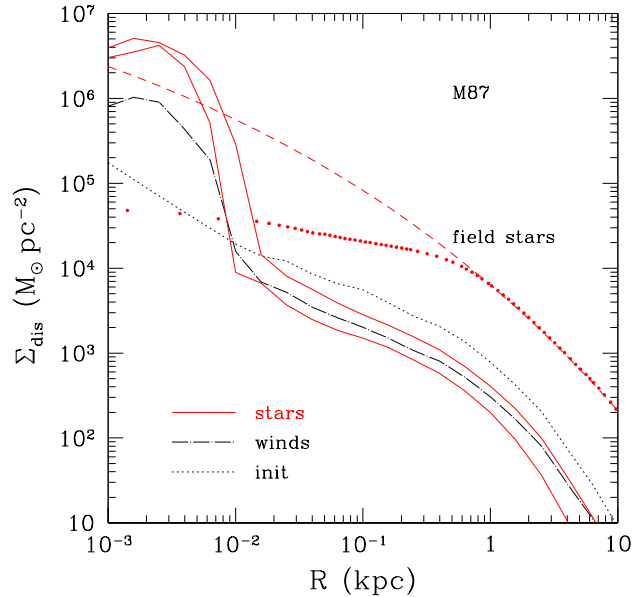}}
\vspace{-0.4cm}
\caption{Projected surface density of stars in the M87 model at $z=0$.
  Dashed red line is the adopted field star profile; the observed
  surface brightness \citep[from][]{kormendy_etal09} flattens in the
  inner $0.7\kpc$ and is shown by filled symbols.  Dotted black line
  is the combined initial profile of all model clusters. Solid red
  lines show the growth of the density of stripped stars: at cosmic
  time $t = 2\Gyr$ (lower line) and at present (upper line).  The
  dot-dashed black line shows the cumulative density of stellar winds
  ejected from the clusters by the present time.  Some of this gas
  could be processed into the {\it in situ} star formation.}
  \label{fig:dend_m87}
\vspace{0.2cm}
\end{figure}

Figure~\ref{fig:dend_m87} shows the build-up of the stellar surface
density with time.  The material lost in stellar winds is deposited
early, and close to the initial location of the clusters, but some gas
is contained within the NSC by continuous mass loss from the
accumulated stars.  This gas may contribute to {\it in situ} star
formation that is required to explain the blue colors of some NSCs.
The sharp upturn of the stellar density at $10\pc$ corresponds roughly
to the radius of the NSC composed of stars from disrupted clusters.  A
significant fraction of the NSC is accumulated already by cosmic time
$t=2\Gyr$ ($z \approx 3.3$).

The observed surface brightness of M87 \citep{kormendy_etal09}
flattens in the inner $0.7\kpc$ relative to our adopted single \Sersic
profile.  Such flattenning is likely to develop because of the
scattering of stars away from the center by merging supermassive black
holes, even if the initial stellar density was higher.  In our models
we do not consider the dynamical effects of supermassive black holes,
focusing instead on the early stages when the build-up of NSC could
produce a seed for the central black hole.  If we used a double
\Sersic profile, with a break at $0.7\kpc$, the accumulated NSC could
be smaller, but it is less clear how the contrast with the field
density would change.  The use of a single \Sersic profile allows us
to include galaxy evolution, described by Equation~(\ref{eq:ms}).  The
data are currently lacking to describe the evolution of a broken
profile, and therefore, in this case the formation history of the
globular cluster system is unconstrained.

\section{Lower-mass elliptical galaxies}
  \label{sec:lowermass}

The stellar debris accumulated in the central few pc is less massive
than the observed black hole in M87, while it is the opposite in the
model of the Galaxy. To explore further the dependence of the NSC mass
on host galaxy properties, we consider two scaled-down versions of the
M87 system.  The first (EG1 model) has stellar mass 4 times lower than
M87, the other (EG2 model) has stellar mass 16 times lower than M87
and similar to the Milky Way.  The other parameters for the two models
are set using the observed galactic scaling relations
\citep[e.g.,][]{guo_etal09}:
    $n_{s} = 4$, 
    $R_{e} = 8.6\kpc$,
    $M_{h} = 5\times 10^{12}\Msun$,
    $r_{s} = 35\kpc$
for EG1, and
    $n_{s} = 2$,
    $R_{e} = 2.5\kpc$,
    $M_{h} = 10^{12}\Msun$,
    $r_{s} = 20\kpc$
for EG2.
These undergo the same evolution with redshift as the M87 model above.
We also keep the same initial fraction of globular clusters and slope
of the initial mass function that we used for the M87 model,
$f_{\mathrm{GC},i} = 0.04$, $\beta=2$.  The maximum cluster mass is
$\Mmax = 2\times 10^{7}\Msun$ for EG1, the same as in our M87 model,
while for EG2 $\Mmax = 10^{7}\Msun$, the same as in our Galaxy model.
Since we do not have observed samples of globular clusters for these
new models, our choice of $\Mmax$ is motivated by similarity of the
galaxy stellar mass.

Table~\ref{tab:sim} shows the surviving cluster fraction and the
accumulated stellar mass in these cases.  The expected
black-hole mass in these two systems is derived from the scaling
relation of \citet{gultekin_etal09}.  Viewed together, our four galaxy
models exhibit a clear trend: the stellar mass in disrupted globular
clusters that can accumulate within a few parsecs of the center is
larger than the observed black hole mass in low-luminosity galaxies,
and smaller in high-luminosity galaxies, with the crossing point at
stellar mass of about $10^{11}\Msun$ in the spheroidal component.

Note that the EG2 model has a significantly more massive nuclear
cluster than the Galaxy model, despite the same total stellar mass.
Only part of the difference is due to the higher initial fraction of
globulars in EG2.  If we reduce the initial cluster fraction in the
EG2 model to that of the MW model (from 0.04 to 0.012), the accreted
mass at 10 pc scales down proportionally by a factor of 3.3.  However,
this does not work the opposite way.  If we increase the initial
fraction of the MW clusters to 0.04, the accreted mass at $10\pc$
increases only by a factor of 1.7.  This asymmetry is due to the
different structure (larger effective radius $R_{e}$) of the late-type
MW model compared to the early-type EG2 model.

\begin{figure}[t]
\vspace{-0.4cm}
\centerline{\epsfxsize=3.5in \epsffile{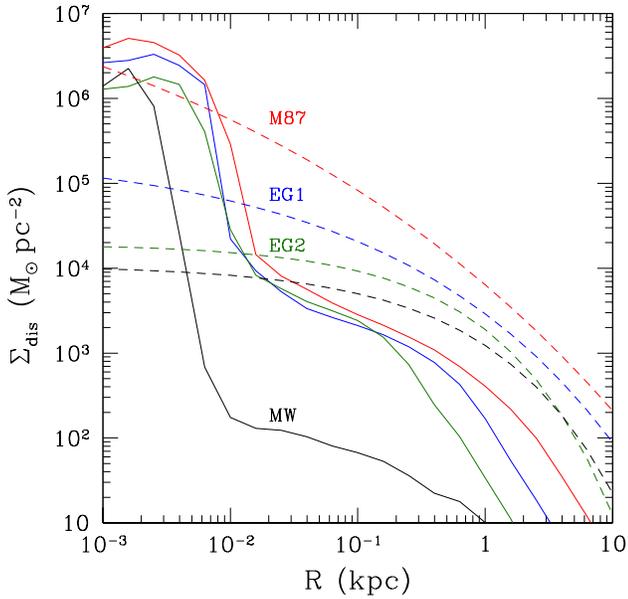}}
\vspace{-0.4cm}
\caption{Current projected surface density of stars in the four galaxy
  models.  Solid lines show the stars stripped from disrupted globular
  clusters, evolved to $z=0$; dashed lines show the adopted field
  stellar profile.}
  \label{fig:dend_all}
\vspace{0.2cm}
\end{figure}

\section{Implications for the Formation of Nuclear Star Clusters and Central Black Holes}
  \label{sec:discussion}

Rapid accumulation of globular cluster debris at the galaxy center
leads to the build-up of a massive and prominent NSC.  In this section
we discuss how the predicted properties of these NSCs compare with the
observations of such clusters in nearby galaxies.

Figure~\ref{fig:dend_all} compares the prominence of the NSC for all
four of our galaxy models.  In all models it rises over the field
stars only in the inner $10\pc$.  However within that region, in all
models but M87 it dominates in surface brightness by one or two orders
of magnitude.  Thus these NSCs could be easily observable given
sufficient spatial resolution.  One potential caveat is that if
mergers of supermassive black holes happen after the NSC formation,
the central density may be significantly reduced.

In the MW model, the projected density at our innermost radius, $1\pc$,
is about $10^{6}\Msun\pc^{-2}$, while the average space density within
$1\pc$ is $2\times 10^{5}\Msun\pc^{-3}$.  Both values are within a
factor of two of the density of the Galactic NSC
\citep{schodel_etal07, oh_etal09}, which we consider to be excellent
agreement given the modeling and observational uncertainties.
However, the model predicts a sharper rise of the NSC density than
observed.  The mass contained within $10\pc$ of the Galactic center is
measured to be $\approx 3\times 10^{7}\Msun$
\citep[e.g.,][]{schodel_etal07, merritt10}, while our model
overestimates it by 50\% (see Table~\ref{tab:sim}).  Our results could
be reconciled with observations if this mass was distributed over a
few tens of parsecs instead, as would be the case if we overestimated
the efficiency of dynamical friction near the Galactic center.  It is
also possible that our adopted model for the field star distribution
near the center underestimates the actual density of stars, which
makes the rise of the NSC density in the inner $4\pc$ more pronounced
than observed.

In the M87 and EG1 models, the maximum surface density is $\sim
4\times 10^{6}\Msun\pc^{-2}$.  At $10\pc$, roughly the extent of the
NSC, the projected density for all three elliptical models is in the
range $\Sigma_{\mathrm{dis}}(10\pc) = 10^{4}-10^{5}\Msun\pc^{-2}$.
These values fall right in the middle of the observed distribution of
surface density (measured at the effective radius) of the most massive
NSCs ($M_{\mathrm{NSC}} = 10^{7}-10^{9}\Msun$) in both nearby
late-type galaxies \citep{hartmann_etal11} and in early-type galaxies
with $M_* < 4\times 10^{10}\Msun$ in the Virgo and Fornax clusters
\citep{turner_etal12}, as well as ultracompact dwarf galaxies
\citep{misgeld_hilker11}.

Our results also compare favorably with the observed fraction of
galaxy stellar mass contained in its NSC.  While the compilations of
\citet{seth_etal08} and \citet{scott_graham13} show a wide variation
from galaxy to galaxy, the median value is $2\times 10^{-3}$ for
early-type spheroidal galaxies, and around $3\times 10^{-4}$ for
spirals.  \citet{turner_etal12} find a mean value of
$(3.6 \pm 0.3)\times 10^{-3}$ for early-type galaxies in the Virgo and
Fornax clusters.  Taking the mass enclosed within $10\pc$ ($M_{10}$)
as an estimate of the NSC mass, our models M87, EG1, and EG2 give the
expected values, $M_{10}/M_{*} = (1-3)\times 10^{-3}$.  However, there
is a trend of the ratio $M_{10}/M_{*}$ decreasing with galaxy mass,
which we discuss in Section~\ref{sec:scaling}.

Based on these comparisons of the mass and density, we conclude that
our models predict the build-up of realistic NSCs from the stars
brought in by infalling globular clusters.  These old stars may not
explain the blue colors of many NSCs which indicate an additional younger
population, but they provide sufficient potential well for the
interstellar gas to condense and form new stars.

\subsection{Mass of the Nuclear Star Cluster}

It may be useful to provide a simple estimate of the accumulated
stellar mass in our model.  As a rough guess, we can assume that all
globular clusters with the initial dynamical friction timescale shorter
than the Hubble time will spiral into the center.  We also note that
the initial cluster distribution mirrors the field stellar
distribution, and therefore we can integrate the galaxy stellar mass
enclosed within a typical radius of inspiral for globular clusters.
Since clusters lose mass along the way by tidal disruption and stellar
evolution, we need to reduce our estimate by a typical factor of 0.25
that corresponds to this mass loss.  The accumulated mass of the NSC
is then
\begin{equation}
  M_{NSC} \approx 0.25\, f_{\mathrm{GC},i}\ M_{*}(<r_{\rm df}).
\end{equation}
We have found empirically that a radius $r_{\rm df} \approx 1.4\kpc$
gives a good estimate for the fiducial models ($\beta=2$, $\Mmax =
(1-2)\times 10^{7}\Msun$), to within a factor of 2 of the full models.
Starting at such a radius in our elliptical galaxy models, a $2\times
10^{6}\Msun$ cluster would spiral into the center in a Hubble time.

\begin{figure}[t]
\vspace{-0.1cm}
\centerline{\epsfxsize=3.5in \epsffile{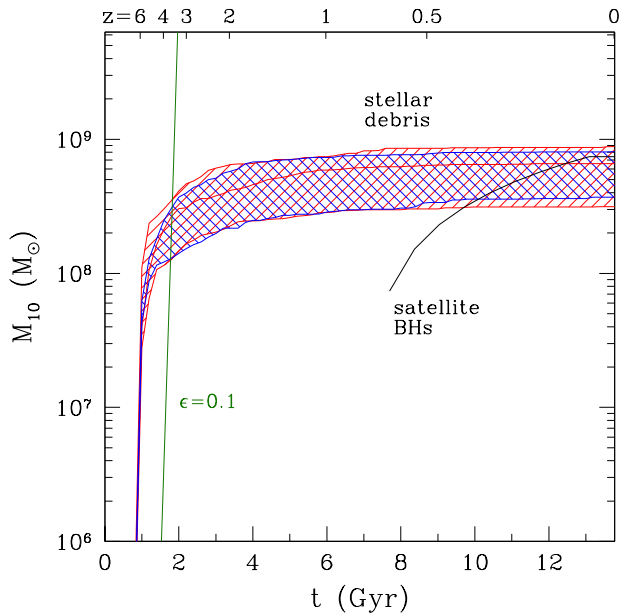}}
\vspace{-0.4cm}
\caption{Stellar mass accumulated within $10\pc$ in the M87 system,
  assuming that the globular clusters begin forming at $z_{i}=6$.
  Line notation as in Figure~\ref{fig:cl_m87}.  The solid black line
  shows the cumulative mass of central black holes from the accreted
  satellite galaxies.  The straight green line indicates the growth of a
  hypothetical $10\Msun$ stellar black hole by continuous
  Eddington-limited accretion with the radiative efficiency
  $\epsilon=0.1$, starting at $z_{i}=6$.}
  \label{fig:m1pc}
\vspace{0.2cm}
\end{figure}

\subsection{Possible Collapse to a Black Hole}
  \label{sec:bh}

Figure~\ref{fig:m1pc} shows the growth with cosmic time of the mass of
stripped stars accumulated within the inner $10\pc$ of M87.  Most of
the mass is assembled by $z=2$, and a significant fraction already by
$z\sim 4$.  This is significantly faster than the growth of the total
stellar mass of the galaxy.

Note that we begin the calculation of cluster evolution at $z_{i}=6$,
and therefore the central mass accumulation begins only from that
epoch.  We did not model earlier epochs because the extrapolation of
the observed galaxy evolution would be extremely uncertain.
Nevertheless, it is likely that the mass began accumulating at an even
higher redshift.  This accumulation would happen on a timescale
shorter than that of radiative accretion, and therefore may have
important implications for the massive black holes powering the
earliest known quasars, as was first suggested by \citet{capuzzo93}.
For illustration we show the growth of a hypothetical $10\Msun$
stellar black hole by continuous Eddington-limited accretion.  We take
a typical value of the radiative efficiency for non-rotating black
holes, $\epsilon=0.1$, which gives the exponential growth (Salpeter)
time $t_{S} = 4.5\times 10^{8}\, \epsilon/(1-\epsilon)\yr = 5\times
10^{7}\yr$.  Such a black hole would reach the mass of $10^{9}\Msun$
in $t_{S} \ln{(10^{9}/10)} \approx 9\times 10^{8}\yr$.  In case of the
maximum radiative efficiency for a rotating black hole, $\epsilon
\approx 0.3-0.4$ \citep{thorne74, noble_etal09}, the timescale is 4--6
times longer.  In either case the initial mass accumulation by
infalling globular clusters is faster.  If this process could create
massive black holes quickly, it would alleviate the timescale problem
posed by the existence of quasars at redshift 7, when the Hubble time
is less than $0.8\Gyr$.

\begin{figure}[t]
\centerline{\epsfxsize=3.5in \epsffile{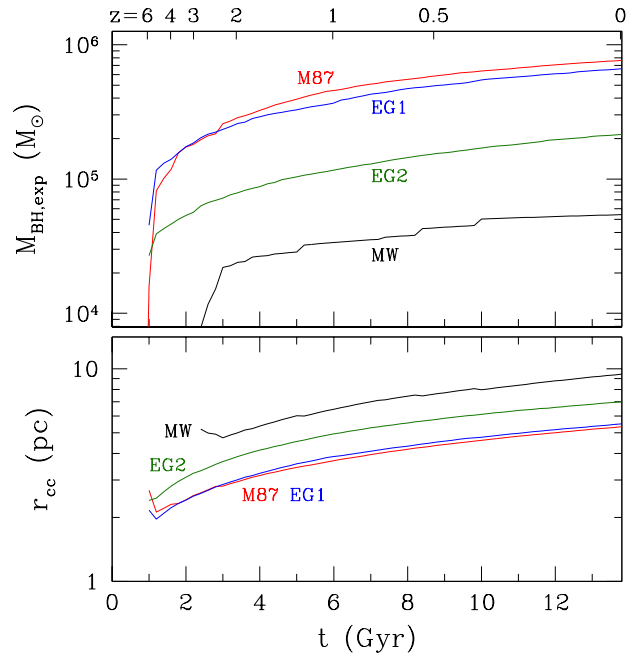}}
\vspace{-0.1cm}
\caption{Evolution of the radius of core collapsed region (bottom
  panel) and the expected mass of a black hole formed within it (top panel),
  according to the estimate given in \S\ref{sec:bh} and
  Equation~(\ref{eq:bhexp}).}
  \label{fig:trel}
\vspace{0.3cm}
\end{figure}

Once a massive NSC forms, its subsequent evolution can indeed produce
a seed black hole at the center.  \citet{miller_davies12} argue that a
cluster with high enough velocity dispersion $\sigma \ga 40\,\kms$
will inevitably form a black hole because kinematic heating from
binary stars is insufficient to prevent complete core collapse.
Collisions between binary and single black holes (products of the
evolution of massive stars) will remove them from the cluster core
after $10^{8} - 10^{9}\yr$, via either three-body kicks or the
asymmetric emission of gravitational radiation.  At the end of this
phase, only one black hole is expected to remain at the center
and grow by tidally disrupting and absorbing other stars in the
collapsing core.  Even if no black hole remains after the violent
kicks, runaway mergers of main-sequence stars will create a single
central black hole.  \citet{miller_davies12} estimate that the
subsequent black hole growth will be fast until the energy of stars
within its sphere of influence is sufficient to hold off collapse.
The expected black hole mass at the end of the consumption process is
given by their Equation~(13), re-written here as
\begin{equation}
  M_{\mathrm{BH,exp}} \sim 700\Msun \left({\sigma \over 40\,\kms}\right)^{4/3} 
     \left({M_{NSC} \over 10^{6}\Msun}\right)^{2/3}.
  \label{eq:bhexp}
\end{equation}
This is a significant boost for growing a supermassive black hole in
galactic nuclei, compared to a typical stellar mass.

There are several uncertainties underlying the above argument.  The
growth of the runaway merger product may be limited by strong stellar
winds, removing the bulk of the mass and leaving a black hole of only
$\sim 10\Msun$ for solar-metallicity stars \citep{glebbeek_etal09}.
However, the extrapolation of the wind model of main-sequence stars
may not apply to the inhomogeneous envelope surrounding merged stellar
cores.  Also, the effect of the winds is reduced at the low
metallicity characteristic of globular cluster stars.  In any case, a
black hole of some substantial mass is expected to form at the
center and to proceed to grow by disrupting and accreting main-sequence
stars.

Another uncertainty concerns the initial phase when stellar-mass black
holes eject each other, until the last one remains to dominate the
core.  If the energy generated in three-body encounters between binary
and single black holes is efficiently transferred to the velocities of
main-sequence stars, core collapse could be delayed by several
relaxation times, until all black holes are gradually ejected
\citep{breen_heggie13b}.  At present, the value of the delay time is
uncertain, as N-body simulations show that it keeps decreasing, by a
factor of several, with increasing range of mass of main-sequence
stars included in the simulation \citep{breen_heggie13a}.  Low
velocity dispersion clusters may indeed retain significant numbers of
black holes for several relaxation times \citep{sippel_hurley13,
  morscher_etal13}, but a fully realistic simulation for a high
dispersion cluster ($\sigma > 40\,\kms$) is not yet
available.  However, as long as one black hole dominates cluster core,
the potential presence of smaller black holes does not change our argument
about runaway growth.

Additional delay of core collapse could be caused by kinematic heating
by the rest of the host galaxy.  We can estimate this delay by
considering the ratio of the nuclear relaxation time to the heating
time, using the scaling given by equation~(10) of \citet{merritt09}:
$t_{\mathrm{rh}}/t_{\mathrm{heat}} \sim (M_*/M_{NSC})^{1/2} \, (R_{NSC}/R_e)^{5/2} 
   \sim 10^{-6}$,
with the typical parameters of NSC and host galaxy from
Table~\ref{tab:sim}.  This ratio is so small that we can ignore the
interaction of NSC with its host galaxy.

In the following discussion we assume that the delay is smaller than
the collapse time of remaining main-sequence stars within the NSC.
The core collapse time depends on cluster concentration and stellar
mass spectrum \citep[e.g.,][]{gnedin_etal99}.  Direct N-body
simulations of \citet{portegies_mcmillan02} and Monte Carlo
simulations of \citet{gurkan_etal04} show that the most massive
component reaches the center within a fraction of the half-mass
relaxation time, typically $t_{cc} \sim 0.2\, t_{\mathrm{rh}}$.  We
use our predicted NSC profile to calculate the relaxation time as a
function of distance from the center, assuming an average stellar mass
of $1\Msun$ and the Coulomb logarithm $\ln{\Lambda}=12$.
The size of the nuclear region $r_{cc}$ that would collapse on a
timescale shorter than the current age, $t_{cc}(r_{cc}) \equiv t$, is
shown in the bottom panel of Figure~\ref{fig:trel}.

Within $\la 1\Gyr$ after the beginning of globular cluster formation,
this region includes a large part of the NSC.  The time to accumulate
half of the final NSC mass ranges from $0.2\Gyr$ for the small EG2
galaxy model to $1.4\Gyr$ for the giant M87 galaxy model.  The core
collapsed region quickly reaches $2-3\pc$ in radius in the early-type
models, where the half-mass radius of the whole NSC is $5-7\pc$
(Table~\ref{tab:sim}).  It is then possible that these dense systems
undergo catastrophic core collapse and produce a seed black hole, as
described by \citet{miller_davies12}.  We estimate the expected black
hole mass using Equation~(\ref{eq:bhexp}) with the estimated parameters
of NSCs in our models.  These masses are plotted in the top panel of
Figure~\ref{fig:trel} and reach $(0.3-1)\times 10^{5}\Msun$ by
$z \approx 5$.

The collisional evolution appears to be slower in the MW model, but
this is an artifact of our starting the calculation at a later epoch
of $z_{i}=3$.  This arbitrary choice was forced by the lack of
information on the early evolution of the Galaxy; in reality the
build-up of the NSC should begin earlier.  The core collapsed region
in the MW model includes essentially the whole NSC, because it is
smaller than in the elliptical models.  The seed black hole mass could
reach $(2-5)\times 10^{4}\Msun$.

\begin{figure}[t]
\vspace{-0.4cm} \centerline{\epsfxsize=3.5in \epsffile{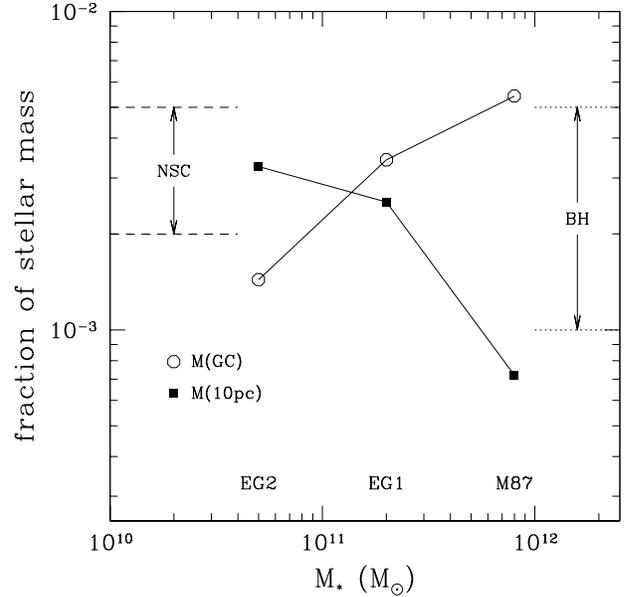}}
\vspace{-0.4cm}
\caption{Scaling with galaxy mass of the fractions of surviving
  globular clusters and nuclear star cluster within $10\pc$, for the
  three early-type galaxy models.  Horizontal dashed lines on the left
  show the median range of NSC mass fraction for galaxies in the local
  neighborhood and the Virgo and Fornax clusters, from
  \citet{seth_etal08, turner_etal12, leigh_etal12, scott_graham13}.
  Horizontal dotted lines on the right show the range of the central
  black hole mass fraction, from \citet{sani_etal11, kormendy_ho13,
    graham_scott13}.}
  \label{fig:sum}
\vspace{0.2cm}
\end{figure}

\subsection{Scaling with Host Galaxy Mass}
  \label{sec:scaling}

Figure~\ref{fig:sum} shows the scaling of our results with galaxy
mass.  The fraction of stellar mass contained in the NSC is a
decreasing function of $M_{*}$ for the three early-type models.
It can be approximately described by the following relation
\begin{equation}
  M_{NSC}/\Msph \approx 0.0025 \, \Msph_{,11}^{-0.5},
  \label{eq:mnsc}
\end{equation}
where $\Msph_{,11}$ is the galaxy stellar mass in units of
$10^{11}\Msun$, and we take $M_{10}$ to be a proxy for $M_{NSC}$.
Using the fundamental plane scaling of galaxy stellar mass with
velocity dispersion, $\sigma_g$, this relation translates to
$M_{NSC} \propto \sigma_g^{1.5-2}$.  It is interesting to note that the MW
model also satisfies this relation if by $\Msph$ we take only the mass
of the bulge, that is, a spheroidal component similar to the
early-type models.

\citet{antonini13} also considered the scaling of NSC mass with galaxy
velocity dispersion and found $M_{NSC} \propto \sigma_g^{3/2}$, in the
model without a pre-existing central BH.  His relation is consistent
with ours and supports the idea that NSC formation is more efficient
in lower-mass galaxies.  If the central BH already exists at the time
when a NSC is built-up, \citet{antonini13} finds that the NSC mass
would be suppressed even further.  This view is supported by
observations of \citet{neumayer_walcher12} that a supermassive
black hole, when present, may limit the NSC growth to $M_{NSC} < 0.01
M_{BH}$.

The observed values of the NSC fraction are scattered over a
large range.  The median for the $M_{*} \la 10^{11}\Msun$ galaxies in
Virgo and Fornax clusters is $M_{NSC}/M_{*} \approx 0.0036$
\citep{turner_etal12}.  The compilations of data for nearby galaxies
by \citet{seth_etal08} and \citet{scott_graham13} both have lower
median values $M_{NSC}/M_{*} \approx 0.002$, and a scaling $M_{NSC}
\propto \sigma_g^{2.1}$.  On the other hand, \citet{leigh_etal12} find
a larger median fraction $M_{NSC}/M_{*} \approx 0.005$ and a steeper
scaling $M_{NSC} \propto \sigma_g^{2.7}$, for 51 early-type galaxies
in lower mass range $M_{*} < 2\times 10^{10}\Msun$.  Given the large
scatter for individual galaxies, on Figure~\ref{fig:sum} we plot only
the median range $M_{NSC}/M_{*} = 0.002-0.005$.

While the NSC fraction decreases with galaxy mass, the fraction of
mass remaining in bound globular clusters instead increases.
$M_{GC}/M_*$ ranges from about 0.001 at $M_{*} < 5\times 10^{10}\Msun$
to 0.005 at $M_{*} \sim 10^{12}\Msun$, consistent with a compilation
of many globular cluster systems by \citet{harris_etal13}, which has
median $M_{GC}/M_* \sim 0.003$, again with large scatter for
individual galaxies.

In addition to considering the mass enclosed within $10\pc$, we also
calculated the whole NSC density profile to the point where it drops
below the density of field stars.  Integrating to $40\pc$ from the
center suffices for all of our models.  We find that the half-mass
radius of the whole NSC scales only weakly with mass, as $R_h \propto
M_{NSC}^{0.23}$.  The observed relation appears to be somewhat
steeper, $R_h \propto M_{NSC}^{0.3-0.5}$ \citep{cote_etal06,
  turner_etal12}, which suggests that our calculations either
overestimate the efficiency of dynamical friction, or underestimate
the efficiency of direct cluster disruption.

\subsection{Why is the Mass of Central Black Hole similar to the Mass of Globular Cluster System?}

One of the initial motivations for this project was to investigate the
puzzling similarity between the mass of the central black hole,
$M_{BH}$, and the combined mass of the globular cluster system in a
given galaxy, $M_{GC}$, discovered by \citet{burkert_tremaine10} in 16
early-type galaxies.  \citet{harris_harris11} and \citet{rhode12}
doubled the sample size and confirmed this result.  It is puzzling
because globular clusters and central black holes currently occupy
very different locations in their host galaxy.

What is the origin of the similarity of $M_{BH}$ and $M_{GC}$?  We
suggest that this correlation is not causal but secondary, resulting
from both quantities being roughly proportional to the total galaxy
mass including dark matter, $M_{h}$.  For black holes it has been well
established that $M_{BH} \propto M_{*}$ \citep[see review
  by][]{kormendy_ho13}, and this relation could also be roughly
expressed through total galaxy mass, as $M_{BH} \sim 10^{-4}\, M_{h}$.
For globular clusters, \citet{spitler_forbes09} and
\citet{harris_etal13} showed that $M_{GC} \approx 6\times 10^{-5}\,
M_{h}$ across four orders of magnitude in galaxy mass (this scaling
was first predicted in theoretical models using cosmological
simulations by \citealt{kravtsov_gnedin05}).  Thus the fractions of
total galaxy mass contained in globular clusters and central black
holes are indeed similar.

The processes of formation and evolution of black holes and globular
clusters are undoubtedly complex.  Is there any physical reason for
the total galaxy mass being the primary, and only, predictor of their
mass?  One of the possible clues is suggested by the success of the
abundance matching technique \citep{kravtsov_klypin99, colin_etal99,
neyrinck_etal04, kravtsov_etal04b, vale_ostriker04} in describing such
varied properties of galaxy population as the redshift- and
scale-dependent clustering strength, average stellar density, and
volume-averaged star formation rate \citep[e.g.,][]{behroozi_etal13}.
In this technique, dark matter halos from cosmological simulations are
matched to real galaxies in a rank-order relation, allowing for random
scatter due to observational uncertainties in the derived luminosity
functions at different redshifts.  One of the main results of such
modeling is the prediction that star formation in galaxies is very
inefficient in general, allowing only a fraction of about 20\% of the
available baryons to be converted to stars in the most efficient
galaxies (with $M_{h} \sim 10^{12}\Msun$) and a much smaller fraction
in the less-efficient galaxies.  Such low fractions indicate that only
central parts of halos are involved in building the observed stellar
systems, including globular clusters and the material that eventually
goes into the central black holes.  The galactic environment (mergers,
tidal interactions, etc.), although important in detail, appears to
play a sub-dominant role to the halo mass.  If this picture is
correct, it explains why galaxy mass controls the properties of such
distinct components as central black holes and extended globular
cluster systems.

However, the growth of $M_{BH}$ and $M_{GC}$ with time may not be
entirely unrelated to each other.  Both could be growing most
efficiently during the episodes of most active star formation, when
the interstellar medium of the host galaxy is very gas-rich and
pressurized.  Such conditions could be realized, for example, during
major mergers of gas-rich galaxies, as suggested for the formation of
globular clusters by \citet{muratov_gnedin10}.

\section{Summary}

The observed profiles of globular cluster systems in normal
galaxies are falling less rapidly with radius than the profiles of the
spheroidal stellar component that has roughly the same age and
metallicity distribution.  This has been interpreted for some time as
indicative of the depletion of GCs in the inner parts of galaxies due
to dynamical processes -- the principal one being dynamical friction,
which drags massive GCs into the nuclear regions.  We quantitatively
estimate these processes, taking the initial GC distributions to be
such that the clusters remaining at the present time fit current epoch
observations.  Thus, we find a good match to current GC properties for
our galaxy and M87, which provides a consistency check for the model.

For the Milky Way the most probable value for the mass of the central
cluster formed by this process is $(2-6)\times 10^{7}\Msun$ and size
roughly $3-5\pc$.  This is close to observational constraints for the
mass and radius of the old stellar component at the Galactic center,
making it probable that this feature of our galaxy was in fact formed
from in-spiraling GCs.  However, our model predicts a larger than
observed density contrast of the NSC relative to the surrounding field
stars, which may be a deficiency of the adopted field profile or the
rate of dynamical friction.

Applying the same method to the giant elliptical galaxy M87, we obtain
a NSC mass of $(3-10)\times 10^{8}\Msun$ and a radius of $5-8\pc$.  In
this calculation we allow for the late-time accretion of satellite
systems, which would also contribute roughly $10^{9}\Msun$ in accreted
black holes to M87, most of which would be expected to merge with the
central BH, adding about 20\% to its mass.  We also construct
evolutionary tracks for two smaller spheroidal stellar systems having
masses of $\Msph = 5\times 10^{10}\Msun$ and $2\times 10^{11}\Msun$,
that produce NSCs of intermediate mass.  Our results for NSC mass
(Table~\ref{tab:sim} and Figure~\ref{fig:sum}) are consistent with the
simple scaling $M_{NSC}/\Msph \approx 0.0025 \, \Msph_{,11}^{-0.5}$.
This result is plausible in that the globular cluster supply is roughly
proportional to the spheroidal component, which would give a constant
ratio, but the efficiency of dynamical friction decreases with the
mass of the system, since the velocities go up and densities go down.
Hence the NSC fraction is a declining function of spheroidal mass.

Adopting a simple prescription that globular cluster formation was
initiated at an epoch $z=6$, we find that the time to assemble the
nuclear clusters is relatively brief, ranging from $0.7\Gyr$ for the
MW to $1.4\Gyr$ for M87.  The stellar relaxation times in these
systems are considerably shorter than the age, so that they would have
sufficient time to reach the core-collapsed state where physical
collisions among main-sequence stars or compact stellar objects are
believed to lead to the formation of central BHs containing a
non-negligible fraction of the cluster mass.  This dynamical process
may lead to the formation of seed BHs of mass $\sim 10^{-3}$ of the NSCs,
which for our galaxy would be in the range $(2-5)\times
10^{4}\Msun$.  For the much more massive systems such as M87, this
process would give a seed mass of $(1-7)\times 10^{5}\Msun$.  While
only a tiny fraction of the final BH, it is amply large to allow
episodic radiative accretion enough time to grow the central BH to its
present mass.  For example, a duty cycle of 4\% (comparable to the
fraction of high-redshift galaxies in AGN mode) with a Salpeter time
of $5\times 10^{7}\yr$ allows a 10,000-fold increase in mass by $z=0$.
We find that the additional increase in mass due to accreted BHs
(embedded in accreted satellite galaxies) is not dominant, but also
not trivial ($1/6$ of the final BH mass), and the late-time mergers
could give a possibly detectable gravitational wave signal
\citep{mcwilliams_etal13}.

It naturally follows from this scenario that essentially all galaxies
with appreciable GC populations should host an old NSC, containing a
fraction $\ga 10^{-3}$ of the spheroidal stellar mass (as given
by equation~\ref{eq:mnsc}).  The crossover galaxy mass, above which
the central BH dominates over the NSC, appears to be at $M_* \sim
10^{11}\Msun$.  In addition to the clear prediction for the prevalence
and properties of NSCs, this scenario also predicts the existence of a
non-trivial number of satellite BHs orbiting massive galaxies, due to
incomplete dynamical friction infall subsequent to minor mergers.

\acknowledgements

We would like to thank the referee for thoughtful comments that helped
to clarify the description of nuclear core collapse, and Cole Miller
for helpful discussions.  We also thank Fabio Antonini, Roberto
Capuzzo-Dolcetta, and Alister Graham for additional comments.
O.Y.G. was supported in part by NASA through grant NNX12AG44G.
S.T. was supported in part by NASA through grant NNX11AF29G.
O.Y.G. thanks the Institute for Advanced Study in Princeton, Kavli
Institute for Theoretical Physics in Santa Barbara, Kavli Institute
for Cosmological Physics in Chicago, and Aspen Center for Physics for
hospitality while this manuscript was being written.

\bibliography{gc}

\end{document}